\documentclass[aps,prd,showpacs,onecolumn,nofootinbib,superscriptaddress,amsmath,amssymb]{revtex4}

\usepackage{subcaption}

\usepackage[normalem]{ulem}

\usepackage{soul}
\usepackage{booktabs}
\usepackage{multirow}
\usepackage{microtype}
\usepackage{graphicx}
\usepackage{graphics}
\usepackage{color}
\usepackage{epsfig}
\usepackage{definitions}%
\usepackage{dsfont,amsmath,bm,braket}

\newcommand{\rZ}[1]{{#1}}

\begin{document}
\title{Hall conductivity of strained $\mathds{Z}_2$ crystals}

\author{I.V. Fialkovsky \footnote{On leave of absence from CMCC-Universidade  Federal  do  ABC,  Santo  Andre,  S.P.,  Brazil}}
\email{ifialk@gmail.com}
\affiliation{Physics Department, Ariel University, Ariel 40700, Israel}

\author{M.A. Zubkov \footnote{On leave of absence from Institute for Theoretical and Experimental Physics, B. Cheremushkinskaya 25, Moscow, 117259, Russia}}
\email{zubkov@itep.ru}
\affiliation{Physics Department, Ariel University, Ariel 40700, Israel}

\date{\today}

\begin{abstract}
We establish topological nature of Hall conductivity of graphene and other  $\mathds{Z}_2$ crystals in 2D and 3D in the presence of inhomogeneous perturbations. To this end the lattice Weyl-Wigner formalism is employed. The non-uniform mechanical stress is considered, along with the spatially varying magnetic field. The relation of the obtained topological invariant to level counting is clarified.
\end{abstract}


\maketitle

\section{Introduction}

\rZ{The $\mathds{Z}_2$ crystals is actually a big family, which includes graphene \cite{Novoselov2005}, the discovery of which won the Nobel Prise.} The study of the latter within Weyl-Wigner formalism is the starting point of our research and its first motivation.

Graphene offers a perfect playground for Quantum Field Theory, both with its lattice methods and algorithms \cite{Hands2008,Drut2009,Buividovich2012,ZW6}, continuous approaches\cite{Fialkovsky:2016kio,Fialkovsky:2011wh}, and comparison thereof. It is due to the combination of lattice nature of the crystal with a very peculiar low energy continuous approximation in \rZ{the} form of quasi-relativistic massless Dirac equation\cite{CastroNeto2009}. The role of the spin is played by the sublattice symmetry (also called pseudo-spin in graphene physics), and its straightforward generalization is $\mathds{Z}_2$ symmetry of crystals. Apart from graphene, the same family includes such 2D materials as  phosphorene, silicene and germanene \cite{Tang2015}. The 3D  $\mathds{Z}_2$  crystals can also be treated within presented formalism, those are, roughly speaking, any of the 14 possible Bravais lattices in 3D if half of the constituent atoms are of a different type.

The aim of this paper is investigation of the topological properties of Hall conductivity of graphene and other 2D and 3D $\mathds{Z}_2$ crystals when \rZ{the} system is subject to a non-uniform external perturbation. To this end, we follow the path developed in \cite{ZW6} and apply \rZ{the} lattice Wigner-Weyl formalism.  The main benefits of this approach is the possibility to reformulate and generalize the  well known topological invariants, like TKNN \cite{TKNN} and others \cite{Volovik2003a,Volovik2017}, to systems in presence of varying external electromagnetic fields, or non-uniform mechanical stress.

Before engaging with our investigation of Hall conductivity of $\mathds{Z}_2$ crystals, we present here the basic formulas of Wigner-Weyl formalism. For self-sufficiency of the paper, all the details of this approach are given in the Appendices \ref{app:WW-form} and \ref{app:WW-QFT}. Also see another paper of this volume \cite{Fialkovsky2020}.

For lattice models we use the following (approximate) definition of the Weyl symbol of an operator
\be
(\hat A)_{W} (x,p )
	= \int_{{\cM}} \D{q} A(p+q /2,p-q /2)  e^{\ii  q  x }.
\label{A_W}
\ee
where $\cM$ is the first Brillouin zone.

Partition function of a non-interacting (or otherwise one-particle) system  can be written in Euclidean space as
\be
Z
	= \int D\bar{\Psi}D\Psi
		\,\, e^{-\tr \(\hat{W}[\Psi,\bar{\Psi}] \(\ii \hat \om -\hat \cH \)\)}
\label{Z01}
\ee
where $\cH$ is its Hamiltonian, and $\hat W$ -- density matrix operator \rZ{with Grassmann - valued matrix elements $\hat W[\Psi,\bar{\Psi}](p,q) =\Psi(p)\bar{\Psi}(q)$.} We shall call $\hat Q\equiv\ii \om -\hat\cH$  \rZ{the} \emph{Dirac operator}.

As derived in Appendix  \ref{app:WW-QFT}, or with somewhat more details in \cite{Fialkovsky2020}, the current density in the system can be given in terms of Weyl symbols of $\hat Q$ and $\hat G=\hat Q^{-1}$ as
\be
	\braket{J_k(x)} = \int dp \, G_W(x,p) \partial_{p_k}Q_W(x,p),
	\label{j(x)}
\ee
\rZ{Averaging over the whole area of the sample, we obtain}
\be
\bar{J_k} \equiv \int \D{x}\braket{J_k(x)} = \Tr( G_W \ast \partial_{p_k} Q_W),
\ee
\rZ{which is the} topological invariant in phase space \cite{Fialkovsky2020}.
Expanding \Ref{j(x)} in powers of field strength $F_{mk}$, we can obtain local conductivity, $\sigma(x)$, see Appendix \ref{app:WW-QFT} for details. Upon calculating its average we come to
\be
\bar\sigma_{mk} =
	\frac{\ii}2 \int dp dx
	\,\tr \(
	G_W^{(0)} \ast \partial_{p_0} Q_W^{(0)} \ast
	G_W^{(0)} \ast \partial_{p_m} Q_W^{(0)} \ast
	G_W^{(0)} \ast \partial_{p_k} Q_W^{(0)}
	\).
	\label{bar_s}
\ee
In $2+1$D the average Hall conductivity  (i.e. the Hall conductivity integrated over the whole area of the sample divided by this area ${\cal A}$) is given by a topological invariant, $\cN$
$$
	\sigma_H = \frac{\cal N}{2\pi},
$$
where
\be
{\cal N}
=  \frac{T \epsilon_{ijk}}{ {\cal A} \,3!\,4\pi^2}\,  \Tr
\[
{G}_W(x,p )\ast \frac{\partial {Q}_W(x,p )}{\partial p_i} \ast \frac{\partial  {G}_W(x,p )}{\partial p_j} \ast \frac{\partial  {Q}_W(x,p )}{\partial p_k}
\]_{A^{(E)}=0}
\label{cN}
\ee
with $\Tr$ defined in \Ref{app:Tr}.

This is a generalization of \rZ{the topological invariant $N_3$ proposed in} \cite{Volovik2003a,Matsuyama1987a} to \rZ{the} inhomogeneous systems. We see, that knowing the Weyl symbol of the Dirac operator of the system, $Q_W$, suffices to analyze the conductivity of the system.

Many nondissipative transport phenomena \rZ{were} already studied within Wigner-Weyl formalism\cite{ZW1,ZW2,ZW3,ZW4,ZW5,ZW6}. \rZ{The} absence of equilibrium chiral magnetic effect \cite{CME} has been proven \cite{ZW5};  anomalous quantum Hall effect studied in the Weyl semimetals and topological insulators \cite{ZW6}, the chiral separation effect \cite{CSE} was investigated  within the lattice models \cite{ZW3,ZW1}. Using the same technique, the investigation of the hypothetical color-flavor locking phase in QCD was performed in \cite{ZW4}, while the scale magnetic effect \cite{SME} has been \rZ{considered} in \cite{ZW2}.

The rest of the paper is organized in following way. In \rZ{the} next Section \ref{sec:Nonlocal} we construct the Hamiltonian and the Dirac operator for a general inhomogeneous system. In Section \ref{sec:elastic} we analyze how the non-uniform elastic deformations enter the above constructed Dirac operator. Finally, in Section \ref{sec:IQHE} we analyze the Hall conductivity under various circumstances. The paper is brought to a close with Conclusion and two Appendices.

\section{Dirac operator for nonlocal tight--binding models and its Weyl symbol}
\label{sec:Nonlocal}
\subsection{Nonuniform Dirac operator}
We begin this Section \rZ{formulating the tight-binding model for the $\mathds{Z}_2$ crystals in $D$ dimensions}, staying within nearest neighbour approximation.
Hamiltonian of a very generic non-uniform model can be written as:
\be
\cH
	=- \sum_{\bx ,\by}\,\, \bar \Psi(\by) t(\by,\bx )
	\Psi(\bx )
\label{bPsi-Psi-gen}
\ee
with the sum running over the lattice sites $\bx, \by\in\cO$, while $t(\bx, \by)$ is matrix of hopping parameters. We suppose that the lattice is infinite, effectively neglecting the finite volume effects.

In a tight-binding approach within the nearest neighbor approximation, permitting the jumps of electrons between the adjacent sites only, the hoping parameters matrix will become
\be
t(\by, \bx)
	= \sum_{j=1}^M \delta(\by- (\bx+\bbj)) t^{(j)}(\by) \label{fg}
\ee
where $\bbj$ are the vectors connecting each atom to its nearest $M$ neighbors, $j = 1,...,M$,  and $t^{(j)}(\by) $ is the non-uniform varying hopping parameter in the direction of $\bbj$. \rZ{$\delta(\bx)$ is the lattice delta function. It is equal to unity for $\bx = 0$ and equal to zero otherwise.}

The Fourier representation is easily constructed then
\be
\begin{split}
t(\bq , \bl)
& = \frac{1}{|{\cM}|}\sum_{j=1}^M \sum_{\bx , \by } e^{-\ii \bq \bx  + \ii  \bl  \by }
	\delta(\by-(\bx+\bbj ))t^{(j)}(\by) \\
&
	= \sum_{j=1}^M  {t}^{(j)}(\bq- \bl) e^{ \ii \bq  \bbj },
\end{split}
\label{f-fourier}
\ee
leading to
\be
\cH
	= \frac{1}{|{\cM}|}
	 \sum_{j=1}^M \int_\cM \D{\bp  d\bq}   \bar \Psi(\bp)
	\[ {t}^{(j)}(\bp-\bq) e^{ \ii \bq  \bbj }\]
	\Psi(\bq).
\ee
The integral here is over the first Brillouin zone $\cM$ specific for the given lattice model. \rZ{$|\cM|$ is its volume.}

In this paper we will limit ourselves with a particular case of crystals exhibiting $\mathds{Z}_2$ symmetry, i.e. when there are two sublattices $\cO_{1,2}$ that \rZ{compose} the crystal, each connected to the other one by one of the $\bbj$ vector
\be
\cO= \cO_1\cup\cO_2,\qquad
	\cO_2=\cO_1+\bbj \equiv \{\bx+\bbj,\quad \bx \in \cO_1\}.
\ee
The choice of the value of $j$ here is actually irrelevant.
The wave function $\Psi(t,\bx)$ can be considered separately on the sites of two sublattices. Moreover, we can define
\be
\Psi_\al(t, \bx)
	=\left\{\begin{array}{ll}
		\Psi(t, \bx),& \bx \in  \cO_\al, \\
		0, & \bx \notin  \cO_\al, \\
	\end{array}\right., \quad \al = 1,2.
\ee
The Hamiltonian is written in a matrix form
\be
\begin{split}
\cH&=\sum_{{\by_1 \in \cO_1}\atop{\by_2 \in \cO_2}}
		\Big(\bar{\Psi}_1(\by_1),\ \bar\Psi_2(\by_2) \Big)
		 {\bm H}(\by_1,\by_2)
		\Big({\Psi}_1(\by_1),\ \Psi_2(\by_2) \Big)^T ,
\end{split}
\label{cH}
\ee
where
\be
{\bm H} =
	\(\begin{array}{cc}
	0 &H_{12}\\
	H_{21} &	0\
	\end{array}\),\qquad
H_{21}(\by_2,\by_1) = -\sum_{j=1}^M \delta\(\by_2-(\by_1+\bbj )\)
	\tj \(\frac{\by_1+\by_2}{2}\),\qquad
	{{\by_1 \in \cO_1}\atop{\by_2 \in \cO_2}}
\label{H21 t}
\ee
and $H_{12}(\by_1,\by_2) =  H_{21}(\by_2,\by_1)$.
For a better readability of the following formulas, we redefined the hoping parameters by their values in the middle of \rZ{the} lattice links, $\tfrac{\by_1+\by_2}{2}$.
	
We shall call $2$ by $2$ matrix operator ${{\bm H}}$ by \emph{lattice Dirac Hamiltonian}, despite that it still describes a generic lattice model, and only upon specifying $\bbj$ its geometry will be established.

Let us consider the off-diagonal term in the Hamiltonian. It can be written in terms of the Fourier transformation as
\be
\cH_{21}=\frac{1}{|{\cM}|}\int_\cM \D{\bp  d\bq }
\bar{\Psi}_2(\bp) H_{21}(\bp,\bq) \Psi_1(\bq)
\ee
Accordingly, the wave functions are
$$
\Psi_\al(\bp) = \frac{1}{|{\cM}|^{1/2}} \sum_{\bx \in \cO_\al} \Psi_\al(\bx)e^{-\ii \bp \bx },
\qquad
\Psi_\al(\bx) =\frac{1}{|{\cM}|^{1/2}} \int_\cM \D{\bp} \Psi_\al(\bp)e^{\ii  \bp \bx }, \qquad\qquad
\al=1,2.
$$
\rZ{The} first Brillouin zone $\cM$ \rZ{is actually the same} for both sublattices as \rZ{they} are constructed \rZ{using} the same basis vectors. 

Modifying Eq. (\ref{f-fourier}) for two sublattices, we have
	\be
	\begin{split}
		H_{1 2}(\bq, \bp)
		& = \frac{1}{|{\cM}|}\sum_{j=1}^M \sum_{\by_1\in\cO_1\atop \by_2\in\cO_2} e^{-\ii \bq \by _1 + \ii \bp  \by _2}
		\delta\(\by_2-(\by_1+\bbj )\)
		\tj \(\frac{\by_1+\by_2}{2}\)\\
		&= \frac{1}{|{\cM}|}
		\sum_{j=1}^M \sum_{\by_1\in\cO_1} e^{-\ii (\bq -\bp ) \by _1 +\ii \bq  \bbj }
		\tj \(\by_1+\bbj /{2}\)
	\end{split}
	\label{f-fourier2}
\ee
Further simplifying this expression we introduce an auxiliary set of points by
\be
	\cO^{(j)}_{{1}/{2}}=\{\by_1+\bbj /{2},\ \by_1\in\cO_1\}.
\ee
Now we can write
\be
	H_{21}(\bp,\bq) =  \sum_{j=1}^M \tj (\bp-\bq) e^{   \ii (\bp + \bq) \bbj /2}
\label{H_12}
\ee
where
\be
\tj(\bp) = \frac{1}{{|{\cM}|}} \sum_{\bx \in \cO^{(j)}_{{1}/{2}}} \tj(\bx) e^{-\ii \bx \bp }
\label{tj}
\ee
The above formula can be interpreted as a Fourier transformation of a function shifted by $\bbj /2$.
When hoping parameters are constant, $t^{(j)}(\bx) = t^{(j)}$, one has
$$
	{t}^{(j)}(\bp) =t^{(j)} \delta(\bp\!\!\!\mod \bgj),
$$
where $\bgj$ are vectors of the reciprocal lattice. This is the expected type of periodicity in momentum space stemming from the topological structure of $\cM$.

The action of the system, and thus the partition function, see \Ref{Z01}, are actually written in terms of the \emph{Dirac operator},
\be
Q\equiv \ii \omega -{\bm H} =
\(\begin{array}{cc}
\ii \omega &-H_{12}\\
- H_{21} &		\ii \omega\\
\end{array}\).
\label{def Q}
\ee
Moreover, the current \Ref{j(x)} and conductivity \Ref{bar_s} are given as functionals of its Weyl symbol, $Q_W$. \rZ{The latter will be considered in the next subsection}.

\subsection{Weyl symbol of Dirac operator}
For calculation of the Weyl symbol of the components of ${\bm H}$ from \Ref{f-fourier} we apply \Ref{A_W} to obtain
\be
H_{21,W} (\bx,\bp )
	= \int_{{\cM}} \D\bq  e^{\ii  \bq  \bx }\sum_{j=1}^M   \tj (\bq)\,  e^{  \ii \bp   \bbj  }
	= \sum_{j=1}^M  e^{\ii   \bp \bbj }   \int_{{\cM}} \D\bq   \tj (\bq) e^{\ii  \bq  \bx }
\ee
For constant hopping parameters \Ref{tj}, it simplifies to
$$
	H_{21,W} (\bx,\bp ) =  e^{\ii   \bp \bbj }  \tj.
$$	
However, in non-uniform systems, one gets
\be
\begin{split}
H_{21,W} (\bx,\bp )
	&=  \frac{1}{|{\cM}|}   \sum_{j=1}^M  e^{\ii   \bp \bbj }  \sum_{\by \in O^{(j)}_{1/2}} \tj(\by)
		\int_{{\cM}} \D\bq  \,    e^{\ii  \bq  (\bx-\by)  }\\
&=  \sum_{j=1}^M  e^{\ii   \bp \bbj }  \sum_{\by \in \cO^{(j)}_{1/2}} \tj(\by) {\cal F}(\bx-\by ),
\end{split}
\label{H21W}
\ee
where
\be
	{\cal F}(\bx)  = \frac{1}{|{\cM}|} \int_{{\cM}} \D\bq  e^{\ii  \bq  \bx  }
\ee
Notice that for $\bx, \by \in \cO^{(j)}_{1/2}$ we have $\by-\bx\in\cO$ and thus, the function ${\cal F}(\by-\bx)$ vanishes for all $ \bx \in \cO^{(j)}_{1/2}$ except for $ \bx =\by$. However, it remains nonzero and oscillates for all other values of $ \bx $, including continuous ones, and gives unity if summed over $\cO^{(j)}_{1/2}$ for any $ \bx $
\be
	\sum_{\by \in \cO^{(j)}_{1/2}} {\cal F}(\bx-\by ) = 1.
\ee
Each term of the $j$-sum in \Ref{H21W} receives \rZ{the} particular form if $ \bx  \in \cO^{(j)}_{1/2}$ (with \rZ{the} same value of $j$):
\be
	H^{(j)}_{21,W} (\bx,\bp )\Big|_{\bx\in \cO^{(j)}_{1/2}} = 	e^{\ii \bp \bbj }  \tj(\bx )
\ee
However, \Ref{H21W} defines $H_W$ also for the continuous values of $ \bx $.

Finally, we shall take into account possible presence of external electromagnetic field with vector potential $ \bA$. To do so, the hopping parameters in $H_{21}$ should be modified
\be
	\tj(\bx) \to \tj(\bx) e^{-\ii \int_{\bx-\bbj /2}^{\bx + \bbj /2}  \bA(\by)d\by },
	\label{pieirls}
\ee
with a complex conjugate substitution in $H_{12}$. From \Ref{H21W} we see that it is simply Pieirls substitution in the language of Weyl symbols.

Combining it with \Ref{pieirls} we obtain
$$
	Q_W = \sum_{j=1}^{M} Q^{(j)}_W
$$
where
\begin{equation}
Q^{(j)}_W(\bx,\bp )\Big|_{\bx \in  \cO^{(j)}_{1/2} }
	= \left(\begin{array}{cc}
		\ii \omega/M & -\tj (\bx)\, e^{\ii  (\bp  \bbj-{A}^{(j)}(\bx))} \\
		-\tj(\bx) \, e^{-i (\bp \bbj-{A}^{(j)}({\bx} ))} & \ii \omega/M
	\end{array}\right)
\label{QWj},
\end{equation}
 $M$ is the number of the nearest neighbours. Here
$$
	A^{(j)}(\bx) = \int_{\bx-\bbj /2}^{\bx + \bbj /2}  \bA(\by)d\by.
$$
For slowly varying hoping parameters $\tj$ and EM potentials $\bA$ one can use Eq. \Ref{QWj} for continuous values of $\bx$ to obtain
\begin{equation}
Q_W(\bx,\bp )=\sum_{j=1}^{M}
	\left(\begin{array}{cc}
		\ii \omega/N & -t^{(j)}(\bx)\, e^{\ii  (\bp  \bbj-{A}^{(j)}(\bx))} \\
		-t^{(j)}(\bx) \, e^{-\ii (\bp \bbj-{A}^{(j)}(\bx))} & \ii \omega/N
	\end{array}\right).
 \label{QW}
\end{equation}
This approximation corresponds to the situation, when the typical wavelength of electromagnetic field is much larger than the lattice spacing.


\section{Elastic deformation and modification of hoping parameters}\label{sec:elastic}

Starting from this section, we will focus on a particular example of a $2+1$D $\mathds{Z}_2$ crystal, the famous mono-layer graphene. Its elastic deformations are very well studied.
Assuming that graphene sheet is parametrized by coordinates $ x_k, k = 1,2$ embedded into three-dimensional space, each point's displacements will have three components $u_a(\bx)$, $a=1,2,3$. The new coordinates are give by:
\begin{eqnarray}
	y_k(\bx)& = & x _k + u_k(\bx), \quad k = 1,2\nonumber\\
	y_3(\bx)& = &u_3(\bx)
\end{eqnarray}
In the absence of the out of plane displacements, i.e. with $u_3=0$, the graphene \rZ{remains} flat. The induced metric of elastic deformation in this case is given in terms of the stress tensor
\be
u_{ik} = \frac{1}{2}\Bigl(\partial_i u_k +
	\partial_k u_i +  \partial_i u_a \partial_k u_a\Bigr),
		\quad a = 1,2,3,
		\quad i,k = 1,2.
\ee
in the following form:
\begin{equation}
g_{ik} = \delta_{ik} + 2 u_{ik}.
	\label{Two_deformations}
\end{equation}
It was further shown in \cite{Volovik2013} that hopping parameters are changed non-trivially by the in-plane strain
\begin{equation}
t^{(j)}(\bx)
	\approx
	t\(1-{\beta} u_{ik}(\bx)  b_i^{(j)} b_k^{(j)}\).
	\label{HoppingElements}
\end{equation}
This approximation is valid under condition that $\beta |u_{ij}| \ll 1$. $\beta$ is called the Gruneisen parameter.
Here we finally put explicitly the graphene geometry by \rZ{pointing out} $\bbj$:
\be
	\{\bbj \}_{j=1}^3 = \{(-1,0); (1/2,\sqrt{3}/2); (1/2,-\sqrt{3}/2)\}.
	\label{l-s-gr}
\ee
The standard expression \cite{Volovik2013} for the emergent electromagnetic potential has the form
\begin{eqnarray}
	{A}_1  & = &- \frac{\beta}{a}\,u_{12}\nonumber\\
	{A}_2  & = & \frac{\beta}{2a}\,(u_{22}-u_{11}) .
\label{AFP2_}
\end{eqnarray}
Combining this with \Ref{QW} we have the following expression for $Q_W$:
\be
Q_W(\omega,p; \tau, \bx ) =
\ii \omega -t  \sum_{j=1}^3
	\( 1- \beta u_{ik}(\bx)  b_i^{(j)} b_k^{(j)} \)
	\(\begin{array}{cc}
		0 & e^{\ii  (\bp  \bbj -{ A}^{(j)}(\bx))} \\
		e^{-\ii (\bp  \bbj -{ A}^{(j)}(\bx ))} & 0
	\end{array}	\),
\label{QW}
\ee
where by $\bA$ we must understand a sum of the induced field \Ref{AFP2_} and any external EM potential present in the system.

An interesting observation is in place if we search for isotropic (although not necessarily uniform) hopping parameters
\be
t^{(1)}(\bx) = t^{(2)}(\bx) = t^{(3)}(\bx).
\label{t-cond}
\ee
With lattice vectors given by \Ref{l-s-gr}, the nontrivial part of $\tj$ is
\be
u_{kl}({\bx})  \bj_k \bj_l =
\frac{a^2}4 \(\begin{array}{c}
4 u_{11}\\
u_{11}+2\sqrt3 u_{12}+3 u_{22}\\
u_{11}-2\sqrt3 u_{12}+3 u_{22}
\end{array}\).
\ee
Then condition \Ref{t-cond}  implies somewhat unexpectedly the Cauchy-Riemann conditions on the pair of displacements $u_1$, $u_2$
$$
\frac{\partial u_1}{\partial x_1} = \frac{\partial u_2}{\partial x_2}, \qquad
\frac{\partial u_2}{\partial x_1}  =-\frac{\partial u_1}{\partial x_2}.
$$
That is, $h(z) \equiv u_1(z) + \ii u_2(z)$ should be analytical function of $z = x_1 +\ii x_2 $. \rZ{Notice}, that there is another solution of \Ref{t-cond}
$$
\partial_1 u_1 = -2-\partial_2 u_2, \quad \partial_1 u_2  = \partial_2 u_1,
$$
which, however, brakes the smallness condition $\beta |u_{ij}|\ll 1$.

\section{Integer Quantum Hall effect in the presence of varying magnetic field and elastic deformations}
\label{sec:IQHE}

Now we can return to investigation of the Hall conductivity, given by \Ref{cN} as $ \sigma_H = \frac{\cal N}{2\pi}$,
with
\be
{\cal N}
=  \frac{T \epsilon_{ijk}}{ {\cal A} \,3!\,4\pi^2}\,  \Tr
\[
{G}_W(x,p )\ast \frac{\partial {Q}_W(x,p )}{\partial p_i} \ast \frac{\partial  {G}_W(x,p )}{\partial p_j} \ast \frac{\partial  {Q}_W(x,p )}{\partial p_k}
\]_{A^{(E)}=0}
\tag{6}
\ee
In \cite{FZ2019} it was shown, that in $2+1$D the topological invariant $\cN$ can be written in the following way:
\begin{eqnarray}
{\cal N} &=&  \frac{\ii\,(2\pi)^2}{8\pi^2\, {\cal A}}\,\sum_{n,k} \int_\dR \D\omega  \epsilon_{ij}\,
\frac{\langle n| [{\cal H}, {\hat x }_i] | k \rangle  \langle k | [{\cal H}, {\hat x }_j] | n \rangle  }
{(\ii\omega^{}-{\cal E}_n)^2 (\ii\omega^{}-{\cal E}_k)}
\nonumber\\
&=&-\frac{2\ii\,(2\pi)^3}{8\pi^2\, {\cal A}}\,\sum_{n,k}   \, \epsilon_{ij}\,
\frac{\theta(-{\cal E}_n)\theta({\cal E}_k)}{({\cal E}_k-{\cal E}_n)^2}
\langle n| [{\cal H}, {\hat x }_i] | k \rangle    \langle k | [{\cal H}, {\hat x }_j] | n \rangle  .
\label{sigmaHH}
\end{eqnarray}
valid for one-particle systems, $\cH\ket{n}=\cE_n\ket{n}$ . This is, of course, nothing else but the conventional expression for the Hall conductance (up to a factor of $2\pi$). It can be reduced to the number of occupied (Landau) levels for systems in external magnetic field (perpendicular to the Hall system) both for constant hoping parameters, and coordinate dependent ones as well.

First of all, let us remind the usual procedure for \rZ{the} former case. It is done \cite{KuboHasegawa1959} by decomposing the coordinates $ x_1$, $x_2$ in relative coordinates $\xi_i$ (with bounded values) and center coordinates $X_i$ (the unbounded part)
\be
\hat{x}_1 = \hat{\xi}_1 + \hat{X}_1,\qquad
	 \hat{x}_2 =   \hat{\xi}_2 + \hat{X}_2,
	 \label{xi-X}
\ee
where
\be
\hat{\xi}_1
	= -\frac{\hat{ p}_2-B  x _1}{B},\qquad
\hat{X}_1
	= \frac{\hat{ p}_2}{ B},\qquad\qquad
\hat{\xi}_2
	= -\frac{\hat{ p}_1}{ B},\qquad
\hat{X}_2
	= \frac{\hat{p}_1- B  x _2}{ B}.
\ee
Then the commutation relations follow:
\be
[\hat{\xi}_1,\hat{\xi}_2]
	=  - [\hat{X}_1,\hat{X}_2]
	= \frac{i}{ B},
\qquad
	[\hat{\xi}_i,\hat{X}_j]=0\quad \forall i,j
	\label{comm-xi-X}
\ee
In Landau gauge, $\bA=(0,B x_1,0)$, the Hamiltonian is a function of $\xi_i$ only, $\cH\equiv \cH(\xi_1,\xi_2)$, then its commutators with $X_j$ simply vanish
\be
	[{\cal H}, \hat{X}_1] =  [{\cal H}, \hat{X}_2] =  0.
\label{[HX]}
\ee
We use these relations to obtain:
\be
\begin{split}
{\cal N} &=  -\frac{2\ii\,(2\pi)^3}{8\pi^2\, {\cal A}}\,\sum_{n,k}
		\frac{1}{({\cal E}_k-{\cal E}_n)^2}
		\braket{n| [{\cal H}, {\hat \xi}_i] | k}
		\braket{k |	[{\cal H}, {\hat \xi}_j] | n}
	\epsilon_{ij} \, \theta(-{\cal E}_n)\theta({\cal E}_k) \\
&=   \frac{2\ii\pi}{{\cal A}}\,\sum_{n,k}  \epsilon_{ij}\,
		\braket{n|  {\hat \xi}_i | k}
		\braket{k |  {\hat \xi}_j | n}
	\theta(-{\cal E}_n)\theta({\cal E}_k).
\end{split}
\label{N-nn0}
\ee
To proceed, one writes $\theta({\cal E}_k)=1-\theta(-{\cal E}_k)$ and the sum containing \rZ{the} second term vanishes due to antisymmetrization. Note, however, that it is only possible for bounded operator $\hat \xi$. Further we write,
\be
\cN
	=   \frac{2\ii\pi}{{\cal A}}\,\sum_{n}
		\, \braket{n|  [{\hat \xi}_1,  {\hat \xi}_2 ]| n}
		 \theta(-{\cal E}_n)
=   -\frac{2\ii\pi}{{\cal A}\, B}\,\sum_{n}   \,
	\braket{n|n} \theta(-{\cal E}_n).
\label{N-nn}
\ee
The eigenstate (multy)index $n$ in Hall systems can be represented as a pair of quantum numbers: momentum component $p_2$, and $m$, which in the ordinary system will correspond to Landau level number,
$$
{\cal H} \ket{n}
	= {\cal H}(\hat{ p}_1, \hat p_2-{  B} \hat x _1)
		\ket{p_2, m}
	= {\cal E}_{m}{(p_2)}
		\ket{p_2, m}, \, m\in Z.
$$
The value of $p_2/B$ corresponds to the center of the semi-classical electron orbit, thus it varies in the interval
$(-L/2, L/2)$ (assuming the system being $L\times L$).
This gives
\be
\cN
	=-\frac{(2\pi)}{{\cal A}}\,\sum_{m}\int_{-L/2}^{L/2} \frac{dp_2 L}{2\pi}  \, \frac{1}{  B}
	\theta(-{\cal E}_m {(p_2)})
	=N  \, {\rm sign}(-{  B}).
  \label{calM2d232}
\ee
Here ${\cal A} = L^2$ is the area of the system while $N$ is the number of occupied branches of the spectrum, we come back to their discussion in the end of the Section.

Now we can turn to \rZ{the} non-homogeneous hoping parameters,
\be
	t^{(j)}(\bx) = t^{(j)}_0 e^{ \bx  \bbf },
	\label{t(x)}
\ee
with some constant spatial vector $ \bbf$. The Dirac Hamiltonian becomes,
\be
{\cal H}(\hat\bx,\hat\bp )
	= \sum_{j=1}^M
	\(
		\begin{array}{cc}-\mu/M & t^{(j)}(\hat\bx)\, e^{\ii  (\hat\bp-\bA(\hat\bx))\bbj} \\
		t^{(j)}(\hat\bx) \, e^{-\ii (\hat\bp-\bA(\hat\bx))\bbj} & -\mu/M \end{array}
	\),
	\label{H_hats}
\ee
here $\mu$ is chemical potential. We notice, that due to the $x$-dependence of $\tj$ the relation \Ref{[HX]} does not hold anymore. However, because of its particular form \Ref{t(x)}, we can note, that (again, in Landau gauge)
\be
	{\cal H} \hat{X}_{j}-\hat{X}_{j} {\cal H}  = \frac{\ii}{2 B} \epsilon^{ji}f_i \, ({\cal H}-\mu),
\ee
implying for $n\ne k$
$$
	\langle n | {\cal H} \hat{X}_{j}-\hat{X}_{j} {\cal H}|k \rangle  = 0
$$
and we come again to \Ref{N-nn0}, and consequently to
\be
{\cal N} =   -\frac{2\pi}{{\cal A}\, B}\,\sum_{n}   \,   \langle n|   n \rangle  \theta(-{\cal E}_n)
	=   -\frac{2\pi}{ B } \,{\rho}
\ee
where now 	$\rho$ is the average density of occupied states.

One can consider further restrictions on coordinate dependence of $t^{(j)}$, and demand that $f_2=0$ in \Ref{t(x)}, then $p_2/B$ \rZ{remains} a good quantum number, and \Ref{calM2d232} is applicable, leading again to the same expression for the (averaged) Hall conductivity
\be
\sigma_H
=  \frac{N}{2\pi}  \, {\rm sign}(-{B}).
  \label{sigma_H}
\ee
One can see, that in the presence of constant magnetic field and the hopping parameters $t^{(j)}= t^{(j)}(x_1)$,  the Hall conductivity is given by the same standard value as for the constant hopping parameters. Since direction $x_1$ is actually arbitrary, related to our ad hoc choice of direction in the Landau gauge, we conclude that for $\tj$ \Ref{t(x)} with any $\bm f$, expression \Ref{sigma_H} holds for Hall conductivity.

Some comments are due in regard to the number of \rZ{the}  occupied levels now, $N$, relevant both to the systems of constant and non-constant $\tj$.
In conventional systems they are counted from the neutrality point, and  this way we come to the standard expression for the Hall conductance of the fermionic system in the presence of constant magnetic and electric fields. On the other hand, in the exact treatment of the honeycomb lattices (e.g., for graphene) the levels are to be counted from the edge of the band. This introduces into the consideration the levels with large negative Chern numbers \cite{Sheng2006}. As a result close to the half filling the conductivity is given by the `Dirac-Landau level index' \cite{Hatsugai2} counted from the zero energy.

However, our approximation is probably valid only up to $|E_F|\sim t$, i.e. in the region between the innermost van Hove singularities \cite{Hatsugai2}. Thus we cannot take correctly into account the above mentioned deep lying levels. On the other hand, for the constant magnetic field in the absence of elastic deformations their contribution is known -- it cancels precisely that of ${\cal N}/(2\pi)$ at the half filling. We denote this term by $\sigma_H^{(0)} = {\cal N}^{(0)}/(2\pi)$, and the final expression for the Hall conductivity becomes
\be
	\sigma_H = \frac{1}{2 \pi}\({\cal N}-{\cal N}^{(0)}\)
	\label{sigmadeformed}
\ee
which we can also rewrite as
\be
	\sigma_H = \frac{N^\prime}{2\pi} {\rm sign}(-{B}),
	\label{sigmadeformed2}
\ee
where $N^\prime$ is counted from the half filling (the LLL being occupied contributes with the factor $1/2$).

This expression is valid for weak elastic deformations which do not modify deep Landau levels, as well as in \rZ{the} presence of slowly varying magnetic field (\rZ{it is assumed that the variations of magnetic field are such that they do not alter the  boundary conditions imposed on the vector potential}). In the latter case, ${\cal N}^{(0)}$ is the value of Eq. (\ref{cN}) at half-filling and it should be calculated \rZ{in the presence of} constant magnetic field and without elastic deformations.

\section{Conclusions}
In the present paper we \rZ{show} how the Wigner-Weyl formalism \rZ{may} be applied to investigation of the (topological) Hall conductivity of graphene and other $\mathds{Z}_2$ crystals under \rZ{the}  inhomogeneous external perturbations. In particular, we \rZ{consider} the non-uniform mechanical strain and varying magnetic field.

The main result of the paper, is \Ref{sigmadeformed} supplied with \Ref{cN} showing that the averaged Hall conductivity for such systems is given by the generalization of the $N_3$ topological invariant to non-uniform systems. It remains constant under weak and small variations of the hoping parameters (i.e. mechanical strain) and/or external magnetic field. While weakness of variations mean that they do drag the system over a topological phase transition, their smallness in this case is to be understood as \rZ{the} requirement that they do not change boundary condition for the system (similar to small gauge transformations).

The Weyl-Wigner formalism applied here is approximate in the case of lattice models, however it is valid for any realistic magnetic fields in Hall systems \cite{Suleymanov2020}. Development of an exact formalism, applicable also to \rZ{the} artificial lattices \cite{Scammell2019}, is the aim of \rZ{the}  ongoing research.

The authors are grateful for sharing ideas, comments and collaboration in the adjacent fields to M.Suleymanov, Xi Wu, and Chunxu Zhang. M.A.Z. is indebted for valuable discussions to G.E.Volovik.










\appendix

\section{Wigner-Weyl formalism}\label{app:WW-form}

Wigner-Weyl formalism consists, speaking informally, in getting rid of all operators and Hilbert spaces in formulating ordinary Quantum Mechanics (QM)\cite{Weyl1927,Wigner1932,Groenewold1946,Moyal1949}.
While it can be understood as a completely independent language for QM, it can also be interpreted as \rZ{the} correspondence between QM operators and functions in phase space,
$$
\hat A\equiv A(\hat x, \hat p)
\quad\leftrightarrow \quad
A_W\equiv A_W(x,p),
$$
such that
\be
(\hat A \hat B)_W = A_W \ast B_W,
\label{app:A*B}
\ee
\be
\tr \hat A = \Tr A_W
\label{app:tr=Tr}
\ee
\be
\Tr(A_W\ast B_W) = \Tr(A_W B_W)
\label{app:no-star}
\ee
with some appropriate definitions for $\ast$--product (associative, non-commutative) and $\Tr$ operation.

Schrodinger equation in this formulation is being replaced by Moyal equation
$$
\frac{\partial \rho}{\partial t} = \frac{H\ast \rho-\rho\ast H }{\ii \hbar}\equiv
\{\!\!\{ H,\rho\}\!\!\}
$$
$\rho$ \rZ{is} the Wigner function, i.e. Weyl symbol of the density matrix. The solution to the Moyal equation can be shown to describe completely any quantum system, see \cite{Zachos2005} and references therein.

In infinite space, Weyl symbol of an operator $\hat A$ can be defined through its matrix elements in momentum space:
\be
A_W(x,p)
= \frac1{(2\pi\hbar)^n}\int d^n q \,e^{\ii q x/\hbar}\braket{p+q/2|\hat A|p-q/2}
\label{app:A_W}
\ee
Moyal product is
\be
\ast = e^{\tfrac{\ii\hbar}2(\cev\partial_x\vec\partial_p-\cev\partial_p\vec\partial_x)}
\label{app:ast-prod}
\ee
Weyl trace operation $\Tr$ stands for \rZ{the} integration over whole phase space and summation over  \rZ{the} inner symmetry indices, if any
\be
\Tr A_W(x,p )\equiv
\int \D{x dp} \tr A_W(x,p ).
\label{app:Tr}
\ee
and \Ref{app:tr=Tr} is easily proved in this case.

\rZ{Definition \Ref{app:A_W} of the Weyl symbol of operator $\hat A$  used  in the description of lattice models becomes}
\begin{eqnarray}
A_W(\bx,\bp ) &=&
\int_{{\cM}}d\bcP e^{ix\bcP}
\bra{\bp+\tfrac\bcP{2}} \hat{A}  \Ket{\bp-\tfrac\bcP{2}}
\end{eqnarray}
The integral over $\bcP$ is over the Brillouin zone ${\cM}$, i.e. in ${\cM}$ we identify the points that differ by  a vector of reciprocal lattice $\bgj$. It was shown \cite{Shlomo} that it is a reasonable approximation for any external EM fields, achievable in a laboratory.

\rZ{Using direct calculations} it can be shown, that Weyl symbol $(AB)_W(\bx,\bp )$ of the product of two operators $\hat A$ and $\hat B$ \rZ{may} be written in terms of \rZ{the} individual Weyl symbols and ordinary  Moyal product \Ref{app:ast-prod},
\begin{equation}\begin{aligned}
&(AB)_W(\bx,\bp )=
\int_{{\cM}} \D{\bcP} \int_{\cM} \D{\bcR}
e^{\ii \bx \bcP}
\Bra{\bp+\tfrac\bcP{2}} \hat{A} \Ket{\bcR}
\Bra{\bcR} \hat{B} \Ket{\bp-\tfrac\bcP{2}}\\
&=\frac{1}{2^D}\int_{{\cM}} \D{\bcP d\bcK }
e^{\ii \bx \bcP}
\Bra{\bp+\tfrac\bcP{2}} \hat{A} \Ket{\bp-\tfrac\bcK{2}}
\Bra{\bp-\tfrac\bcK{2}}\hat{B} \Ket{\bp-\tfrac\bcP{2}}\\
&= \[ \int_{{\cM}} \D\bq  e^{\ii \bx  \bq}
\Bra{\bp+\tfrac\bq {2}} \hat{A} \Ket{\bp-\tfrac\bq {2}}
\]
e^{\tfrac{\ii}{2} \(- \cev{\partial}_\bp\vec{\partial}_\bx+\cev{\partial}_{\bx}\vec{\partial}_\bp\)}
\[ \int_{{\cM}} \D\bk e^{\ii \bx   \bk}
\Bra{\bp+\tfrac{ \bk}{2}}\hat{B} \Ket{\bp-\tfrac{ \bk}{2}}
\]
\label{app:Z}
\end{aligned}
\end{equation}
In the second line we change variables
$$
\bcP = \bq+ \bk , \quad \bcK = \bq- \bk
$$
with the Jacobian $ J = 2^D $, which cancels the factor coming from the change of the integration area. Here $D$ is the dimension of space. As usually, the bra- and ket- vectors in momentum space are defined modulo vectors of reciprocal lattice $\bgj$,  as it is inflicted by the periodicity of the lattice.

Notice, that for the chosen form of Wigner transformation on \rZ{the} lattice the above equality is approximate and works only if the operators $\hat{A}$, $\hat{B}$ are close to  diagonal, i.e.
such that their matrix elements $\Bra{\bp+\frac\bq {2}} \hat{A} \Ket{\bp-\frac\bq {2}}$ and $\Bra{\bp+\frac\bq {2}} \hat{B} \Ket{\bp-\frac\bq {2}}$ are nonzero only when $\bq$ remains in the small vicinity of zero.

An important consequence of the formalism is the Groenewold equation relating the Weyl symbol of the Dirac operator and its Green's function. \rZ{On} the operator lever they are simply inverse,
\be
\hat Q \hat G = {\mathds 1},
\ee
then calculating Weyl symbol of both sides, we obtain
\be
(\hat Q \hat G)_W
= Q_W\ast G_W = 1.
\label{app:GreoEqu}
\ee
This equation can be solved iteratively, for \rZ{the} detailed treatment see \cite{Shlomo}. For the \rZ{purposes  of the present paper} we will only need the obvious first approximation
\be
G_W\approx G_W^{(0)} - G_W^{(0)}*Q_W^{(1)}*G_W^{(0)},
\ee
valid for $ Q_W\approx Q_W^{(0)}+ Q_W^{(1)}$.

\section{Wigner-Weyl field theory}
\label{app:WW-QFT}
Partition function of a general model can be written in Euclidian space as
\rZ{\be
Z = \int D\bar{\Psi}D\Psi
\,\, e^{S[\Psi,\bar{\Psi}]}
\label{app:Z01}
\ee}
with the action
\be
S[\Psi,\bar{\Psi}]= \int \D{pdq}
\bar{\Psi}^T(p)\,{Q}(p,q)\,\Psi(q),
\ee
where the integration volume and normalization are to be chosen appropriately for \rZ{the} model under consideration.

As usually, we relate operators $\hat{Q}$ and its inverse, the Green function $\hat{G} = \hat{Q}^{-1}$, acting in Hilbert space ${\cal H}$ with their matrix elements $Q(p,q)$ and ${G}(p,q)$
$$
Q(p,q) = \braket{p|\hat{Q}| q}, \quad
{G}(p,q) = \braket{p|\hat{Q}^{-1}| q}.
$$
It is implied that the basis of $\cal H$ is normalized as $\langle p| q\rangle = \delta(p_{D+1}-q_{D+1})\delta^{(D)}(\bp-\bq)$. The mentioned operators satisfy
\be
	\hat{Q} \hat{G} = {\mathds 1}.
	\label{app:QG=1}
\ee

Now we observe that the action can be represented as a trace of a product of operators
\be
S[\Psi,\bar{\Psi}] = \tr \(\hat{W}[\Psi,\bar{\Psi}] \hat Q\),
\ee
where the Wigner operator is
\be
\hat W[\Psi,\bar{\Psi}] =\ket\Psi\bra\Psi.
\ee
Now variation of partition function is
\rZ{\be
\delta Z = \int D\bar{\Psi}D\Psi \, e^{S }
\, \tr\(\hat{W} \delta \hat Q\)
= Z \tr\(\braket{\hat W} \delta \hat Q\)\\
\ee}
where the usual vacuum EV was used,
\be
\langle \hat O \rangle = \frac1Z\int D\bar{\Psi}D \Psi \, \hat O e^{S[\Psi,\bar{\Psi}]}.
\ee
Further \rZ{applying} Peierls substitution, i.e. noting that \rZ{the} introduction of \rZ{the} EM potential $A$ is simply shifting the momenta, $\bp\to\bp-\bA(x)$,  we obtain
\be
\delta \hat Q = - \partial_{p_k}\hat Q \delta A_k
\ee
and using the basic Weyl transformation properties \Ref{app:tr=Tr} and \Ref{app:no-star}, we come to
\be
\begin{split}
	\delta Z &	=  \rZ{Z}\int dp dx \,\tr\( G_W(x,p) * \partial_{p_k}Q_W(x,p) \delta A(x)\)\\
	&
	=  \int dx \delta A(x) \int dp\, G_W(x,p) \partial_{p_k}Q_W(x,p).
	\nonumber
\end{split}
\ee
Thus the current density is
\be
\braket{J_k(x)} = \rZ{-}\int dp \, G_W(x,p) \partial_{p_k}Q_W(x,p).
\label{app:j(x)}
\ee
Note that it is not a topological invariant, it must averaged over the whole of the sample to have this property. Indeed, the total current is topological invariant
\be
\bar{J_k} \equiv \int \D{x}\braket{J_k(x)} = \rZ{-}\Tr( G_W \ast \partial_{p_k} Q_W).
\ee
Under small variations of the Dirac operator Weyl symbol, $ Q_W\approx Q_W^{(0)}+ Q_W^{(1)}$, the Green's function varies accordingly, $G_W\approx G_W^{(0)}+   G_{W}^{(1)}$, and then
$$
\delta\( \Tr\[G_W \ast \partial_{p_k} Q_W \] \)
=  \Tr\[G_W^{(0)}\ast  \partial_{p_k} Q_W^{(1)}+G_{W}^{(1)}\ast\partial_{p_k} Q_W^{(0)} \]	
$$
Given that $ G_{W}^{(1)} = -G^{(0)}\ast Q_W^{(1)} \ast G^{(0)} $, which follows from Eq. \Ref{app:GreoEqu}, the latter two terms become
\be
\begin{split}
	\Tr&\[ G_W^{(0)}\ast  \partial_{p_k} Q_W^{(1)}-G^{(0)}*Q^{(1)}_W*G^{(0)}
	\ast\partial_{p_k}  Q_W^{(0)} \] \\
	& = \Tr\[ G^{(0)}\ast \partial_{p_k} Q_W^{(0)} \ast G^{(0)}\ast Q_W^{(1)}-G^{(0)}
	\ast Q^{(1)}_W\ast G^{(0)}\ast\partial_{p_k} Q_W^{(0)} \] 	\nonumber\\	
\end{split}
\ee
where we integrated by parts and used that $ \partial_{p_l}G_{W}^{(0)} = -G^{(0)}\ast \partial_{p_l} Q_W^{(0)} \ast G^{(0)} $. Now simple cycling inside the trace proves that
\be
\delta \bar{J_k} =0.
\label{app:deltaJ}
\ee

To obtain the conductivity let us expand the current density \Ref{app:j(x)} in powers of $A$ and its derivatives, assuming that it is actually given as a sum of the two contributions:
$$
A = A^{(M)} + A^{(E)}
$$
where $A^{(E)}$ is responsible for the electric field while $A^{(M)}$ produces the magnetic one. Provided the former one is  weak and slowly varying we shall have
\be
\braket{J(x)}
\equiv  \int dp \, G_W(x,p) \partial_{p_k}Q_W(x,p)
\approx  j^{(0)}+j^{(1)}_l   A^l+
j^{(2)}_{lm}
F^{lm}+\ldots
\label{app:Jappr}
\ee
The first term here \rZ{is expected to be zero for the wide ranges of the systems in accordance to the Bloch theorem} on spontaneous currents in the ground states \cite{ZZ2019_3}, while the second \rZ{term} should be absent \rZ{due to the }  gauge invariance.

To obtain $j^{(2)}_{lm} $, which eventually defines the conductivity, \rZ{we represent
\be
Q_W \approx  Q_W^{(0)}-  \partial_{p_m} Q_W^{(0)} A^{(E)}_m
\label{app:Qappr}
\ee}
The Groenewold equation \Ref{app:GreoEqu} connecting $G_W$ and $Q_W$
$$
G_W * Q_W = 1
$$
can be solved iteratively producing
\be
G_W\approx G_W^{(0)} + G_W^{(0)}\ast(\partial_{p_m} Q_W^{(0)} \rZ{A^{(E)}_m}) \ast G_W^{(0)}.
\label{app:Gappr}
\ee
Further expanding the stars, which contains derivatives in $x$ acting on $A$, we have
$$
G_W\approx G_W^{(0)}+  G_{W,m}^{(1)} A_m + 	G_{W,lm}^{(2)} \partial_l \rZ{A^{(E)}_m},	$$
where
$$
G_{W,m}^{(1)} = G_W^{(0)}\ast \partial_{p_m} Q_W^{(0)} \ast G_W^{(0)},\qquad
G_{W,lm}^{(2)} = \frac\ii2 G_W^{(0)}\ast \partial_{p_l} Q_W^{(0)} \ast   G_W^{(0)}
\ast \partial_{p_m} Q_W^{(0)} \ast   G_W^{(0)}
$$
Upon substitution of \Ref{app:Qappr} and \Ref{app:Gappr} into \Ref{app:Jappr} we obtain
\be
\braket{J_k(x)}
\approx \frac{\ii  \rZ{F^{(E)}_{lm }}(x)}2 \int dp
\,\tr \(
G_W^{(0)} \ast \partial_{p_l} Q_W^{(0)} \ast
G_W^{(0)} \ast \partial_{p_m} Q_W^{(0)} \ast
G_W^{(0)} \boldsymbol{\cdot} \partial_{p_k} Q_W^{(0)}
\)
\ee
where instead of the \rZ{star in the last product we insert} an ordinary one. Note that $Q_W^{(0)}=Q_W^{(0)}(x,p)$,  $G_W^{(0)}=G_W^{(0)}(x,p)$.

Thus, average conductivity (proportional to conductance) is given by
\be
\bar\sigma_{mk} =
\frac{\ii}2 \int dp dx
\,\tr \(
G_W^{(0)} \ast \partial_{p_0} Q_W^{(0)} \ast
G_W^{(0)} \ast \partial_{p_m} Q_W^{(0)} \ast
G_W^{(0)} \ast \partial_{p_k} Q_W^{(0)}
\)
\label{app:bar_s}
\ee
where we restored the $\ast$--product in the last factor using once again \Ref{app:no-star}. From now on we will omit the superscript $(0)$ for \rZ{brevity}.

In \rZ{the} two dimensional case, for \rZ{the} system in \rZ{the} presence of constant magnetic field we come to the following representation of Hall conductivity  averaged over the whole area of the sample ${\cal A}$
$$
\sigma_H = \frac{\cal N}{2\pi}.
$$
Here
\be
{\cal N}
=  \frac{T \epsilon_{ijk}}{ {\cal A} \,3!\,4\pi^2}\,  \Tr
\[
{G}_W(x,p )\ast \frac{\partial {Q}_W(x,p )}{\partial p_i} \ast \frac{\partial  {G}_W(x,p )}{\partial p_j} \ast \frac{\partial  {Q}_W(x,p )}{\partial p_k}
\]_{A^{(E)}=0}
\label{app:calM2d230}
\ee
with $\Tr$ defined in \Ref{app:Tr}.  $\cal N$ \rZ{is} given by \Ref{app:calM2d230} or, equivalently, $\bar\sigma$ of \Ref{app:bar_s} is \rZ{the} topological invariant in phase space, as it can be readily checked \rZ{similar} to \Ref{app:deltaJ}.

\bibliography{wigner2,wigner3}

\end{document}